\documentclass[traditabstract]{aa}
\bibpunct{(}{)}{;}{a}{}{,} 
\usepackage{graphicx}
\usepackage[varg]{txfonts}
\usepackage{color}
\usepackage[switch,pagewise]{lineno}

\begin{document}

\title{Spectropolarimetry of the changing-look active galactic nucleus Mrk~1018\thanks{Based on observations made with the ESO VLT at the Paranal Observatory under program ID~104.B-0620.}}
\author{D. Hutsem\'ekers\inst{1,}\thanks{Research Director F.R.S.-FNRS},
        B. Ag{\'\i}s Gonz\'alez\inst{1},
        F. Marin\inst{2},
        D. Sluse\inst{1}
        }
\institute{
    Institut d'Astrophysique et de G\'eophysique, Universit\'e de Li\`ege, All\'ee du 6 Ao\^ut 19c, 4000 Li\`ege, Belgium
    \and
    Universit\'e de Strasbourg, CNRS, Observatoire Astronomique de Strasbourg, UMR 7550, F-67000 Strasbourg, France
    }
\date{Received ; accepted: }
\titlerunning{Spectropolarimetry of Mrk1018} 
\authorrunning{D. Hutsem\'ekers et al.}
\abstract{We have obtained new spectropolarimetric observations at visible wavelengths of the changing-look active galactic nucleus (AGN) Mrk~1018. The AGN direct spectrum shows an extremely weak continuum with faint broad H$\beta$ and H$\alpha$ emission lines. Both lines can be fit with a single very broad emission line component of full width at half maximum FWHM $\simeq$~7200~km~s$^{-1}$, with no evidence of the additional 3000~km~s$^{-1}$-wide component that was previously detected. While this is in agreement with line formation in a Keplerian disk, the line profile variability suggests that the broad emission line region is likely more complex. The continuum polarization of Mrk~1018 is low; it is not higher in the current faint state compared to the past bright state, confirming that dust obscuration is not the mechanism at the origin of the change of look. The polarization profile of the H$\alpha$ line is asymmetric with no rotation of the polarization angle, which possibly reveals line formation in a polar outflow. Alternatively, the polarization profile may be the consequence of a time delay between the direct and the polarized light. Interestingly, the polarization signatures predicted for broad lines emitted around supermassive binary black holes are not observed.
}
\maketitle
%
%
%

\section{Introduction}
\label{sec:intro}

Mrk~1018 (redshift $z$ = 0.043) is an active galactic nucleus (AGN) that changed from Seyfert type~1.9 to type~1 between 1979 and 1984 \citep{1986Cohen}, then back to type~1.9 in 2015 \citep{2016McElroy}. After a rapid decline of two magnitudes, the AGN brightness reached a minimum in October 2016, showing thereafter low level variability \citep{2017Krumpe}. These changes were explained by variations in the rate of accretion onto a supermassive black hole \citep{2016Husemann}. Such large spectral variations on short timescales (years to decades) characterize changing-look AGNs. This relatively rare phenomenon occurs in both Seyfert galaxies and quasars, challenging our understanding of accretion processes \citep{1976Tohline, 1984Penston, 1985Alloin, 1986Cohen, 2016McElroy, 2015LaMassa, 2016MacLeod, 2019Ross}.

The measurement of the polarization of changing-look AGNs can constrain the mechanism at the origin of their variations, while the time delay expected between the direct and the polarized light can be used to map the scattering regions \citep{2017Hutsemekers,2017Marin,2019Hutsemekers, 2020Marin}. The polarization of Mrk~1018 was only measured once on February 9, 1986, by \citet{1989Goodrich}, who reported a linear polarization degree $p$ = 0.28$\pm$0.05\% and a polarization position angle $\theta$ = 165$\pm$5\degr\ in the 4180$-$6903~\AA\ optical band. In 1986, Mrk~1018 was in a bright type~1 state. We report on new spectropolarimetric observations secured in 2019 when Mrk~1018 was in a faint state. Observations and data reduction are described in Sect.~\ref{sec:obs}. The characteristics of the direct spectrum and the polarization properties are presented and discussed in Sect.~\ref{sec:results}. Conclusions are given in Sect.~\ref{sec:conclu}.

\section{Observations and data reduction}
\label{sec:obs}

Spectropolarimetric observations of Mrk~1018 were carried out on October 26, 2019, using the European Southern Observatory (ESO) Very Large Telescope (VLT) equipped with the focal reducer and low dispersion spectrograph FORS2 that is mounted at the Cassegrain focus of Unit Telescope \#1 (Antu). Linear spectropolarimetry was performed by inserting a Wollaston prism into the beam that splits the incoming light rays into two orthogonally polarized beams separated by 22$\arcsec$ on the detector. Spectra were secured with the grism 300V and the order sorting filter GG435 so as to cover the spectral range 4450$-$8750~\AA\ with an average resolving power R $\simeq$ 440. Because the two orthogonally polarized images of the object are recorded simultaneously, the polarization measurements are essentially independent of variable atmospheric transparency and seeing. In order to derive the normalized Stokes parameters $q(\lambda)$ and $u(\lambda)$, four frames were obtained with the halfwave plate rotated at four different position angles, 0\degr, 22.5\degr, 45\degr, and 67.5\degr. This combination allows us to remove most of the instrumental polarization. The full sequence, constituting an observing block, was repeated three times. The multi-object spectroscopy (MOS) slit was 1\arcsec~wide and 20\arcsec~long, with a spatial scale of 0$\farcs$25 per pixel. The seeing during the observations was around 1\arcsec. 

\begin{figure}[t]
\centering
\resizebox{\hsize}{!}{\includegraphics*{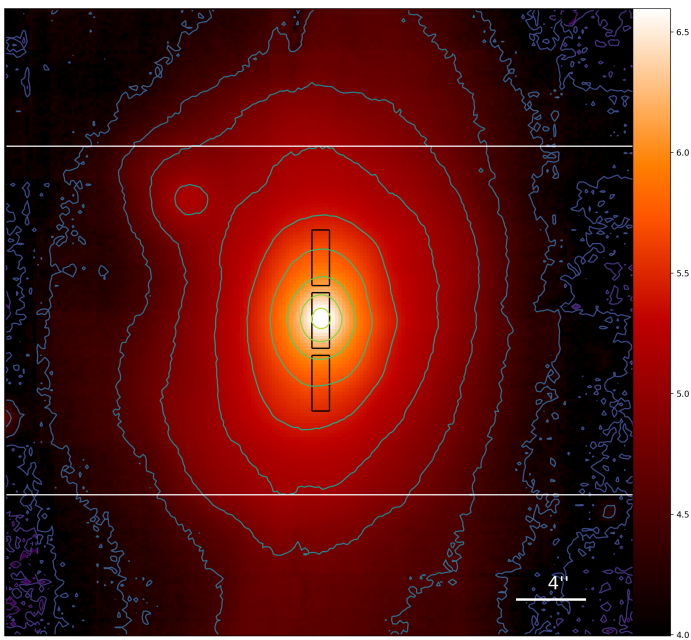}}
\caption{Visible image of Mrk~1018 obtained with the MUSE integral spectrograph \citep{2016McElroy}. North is up and east is left. The intensity scale is logarithmic (arbitrary units). The positions of the 1\arcsec $\times$ 3$\farcs$25 subslits used for the spectrum extraction are indicated. The horizontal lines delineate the MOS strip.}
\label{fig:ima}
\end{figure}

Raw frames were first processed to remove cosmic ray hits using the python implementation of the ``lacosmic'' package \citep{2001VanDokkum,2012VanDokkum}. The ESO FORS2 pipeline \citep{2019vltman} was then used to get images with two-dimensional spectra that are rectified and calibrated in wavelength. The one-dimensional spectra were extracted using a 3$\farcs$25 (13 pixels) -long subslit centered on the nucleus (about three times the seeing value). Since the object extends over the full MOS slit, the sky spectrum was estimated from adjacent MOS strips and then subtracted from the nucleus spectrum. In total, 24 spectra (from two orthogonal polarizations, four halfwave plate angles, and three observing blocks) were obtained. The spectra from the three observing blocks were averaged. The normalized Stokes parameters $q(\lambda)$ and $u(\lambda)$, the linear polarization degree $p(\lambda)$, the polarization position angle $\theta(\lambda)$, and the total flux density $F(\lambda)$ were then computed from the individual spectra according to standard formulae \citep[e.g.,][]{2018Hutsemekers}.  The direct spectrum $F(\lambda)$ was calibrated in flux using a master response curve that does not include the effect of the polarization optics ($q(\lambda)$, $u(\lambda)$, $p(\lambda),$ and $\theta(\lambda)$, on the other hand, are independent of the flux calibration). In order to express $\theta(\lambda)$ with respect to the north-south direction, the spectra were corrected for the rotator angle and for the retarder plate zero angle that is provided in the FORS2 manual \citep{2019vltman}. In addition to the spectra extracted at the position of the nucleus, we also extracted two host galaxy spectra from north and south of the nucleus, as illustrated in Fig.~\ref{fig:ima}. Polarized (BD$-$12~5133) and unpolarized (WD~2039$-$202) standard stars \citep{2007Fossati} were observed and reduced in the same way to check the whole reduction process. WD~2039$-$202 was used to correct the spectra of Mrk~1018 for the telluric absorptions, in particular the O$_{\rm 2}$ B-band that affects the red wing of the H$\alpha$ emission line. 

Spectra of the nucleus and the host galaxy were extracted, using the same subslits, from the data obtained in 2015 with the VLT integral field spectrograph MUSE\footnote{Data obtained from the ESO Science Archive Facility under program ID 094.B-0345.} and from which the recent change of look was identified \citep{2016McElroy}. The ratio of the FORS2 and MUSE spectra was found to slightly and linearly depend on wavelength. Since the same dependence was found for both the nucleus and the host galaxy spectra, it has been attributed to the imperfect flux calibration of the FORS2 spectra, which were corrected accordingly.

\section{Analysis and results}
\label{sec:results}

\subsection{The direct spectrum}

\begin{figure}[t]
\centering
\resizebox{\hsize}{!}{\includegraphics*{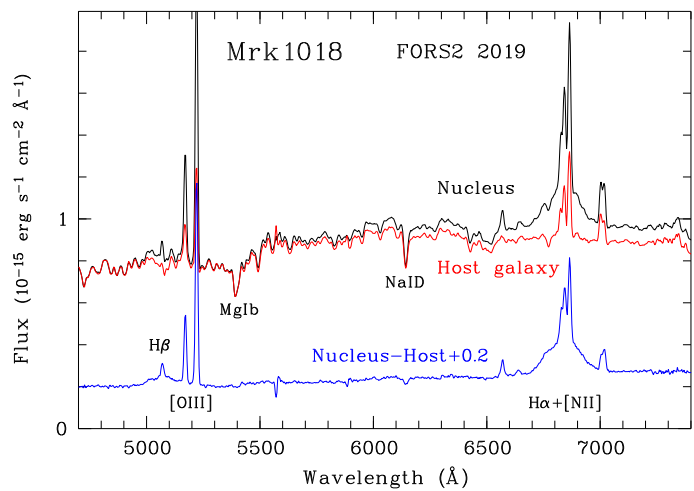}}
\caption{Spectrum of Mrk~1018 covering the H$\beta$ - H$\alpha$ spectral range, observed with FORS2 in October 2019. The nucleus spectrum is shown in black, the underlying host galaxy spectrum in red, and the difference spectrum in blue.}
\label{fig:spec1}
\end{figure}

The FORS2 spectrum of Mrk~1018 is illustrated in Fig.~\ref{fig:spec1}. The nucleus shows several narrow emission lines together with a broad H$\alpha$ emission line. The absorption lines reveal a significant contribution from the host galaxy. To remove the host galaxy contribution, we built a template using the two off-nucleus spectra (Fig.~\ref{fig:ima}). These spectra were averaged after rebinning in wavelength to account for their radial velocity with respect to the nucleus, which is $-$90 km~s$^{-1}$ for the southern part and $+$90 km~s$^{-1}$ for the northern one. The host galaxy spectrum was then scaled and subtracted from the nucleus spectrum. The scaling factor was chosen to minimize the contribution of the MgI$\,$b and NaI$\,$D absorption features in the difference spectrum, keeping it positive in the spectral range of interest.  The resulting difference spectrum isolates the AGN spectrum that shows almost no remaining continuum (hereafter we refer to the difference spectrum as the AGN spectrum and to the total spectrum from the central subslit as the nucleus spectrum).  Mrk~1018 was still in a faint state as of October 2019. Broad H$\beta$ is clearly detected, suggesting a type~1.5 Seyfert classification\footnote{\citet{2016McElroy}, on the other hand, proposed a type~1.9 classification from the comparison of the total spectrum with the original type~1.9 spectrum described by \citet{1981Osterbrock}, although they reported faint broad H$\beta$ emission in their 2015 spectra after continuum subtraction. This outlines the ambiguity of the subtype definition, which depends on the considered spectrum.}.

\begin{figure}[h]
\centering
\resizebox{\hsize}{!}{\includegraphics*{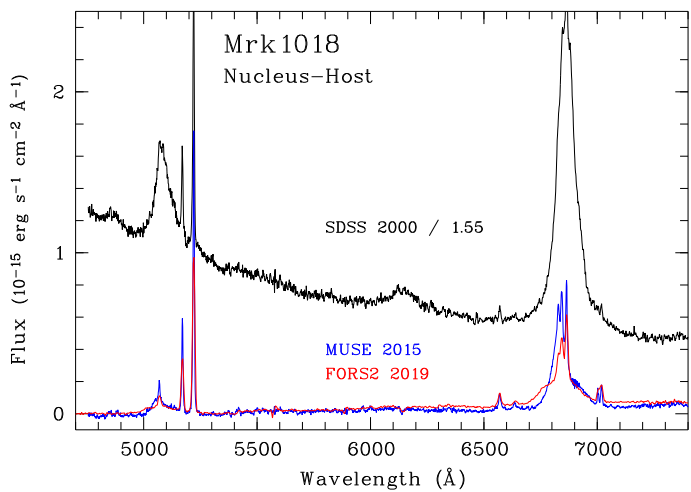}}
\caption{Spectrum of Mrk~1018 at different epochs once the contribution of the host galaxy has been subtracted from the spectrum of the nucleus. The flux density of the SDSS spectrum has been divided by a factor of 1.55 for comparison with the other spectra, which are affected differently by slit loss. Narrow lines appear broader in the FORS2 spectrum due to the lower spectral resolution.}
\label{fig:spec2}
\end{figure}

\begin{figure}[h]
\centering
\resizebox{\hsize}{!}{\includegraphics*{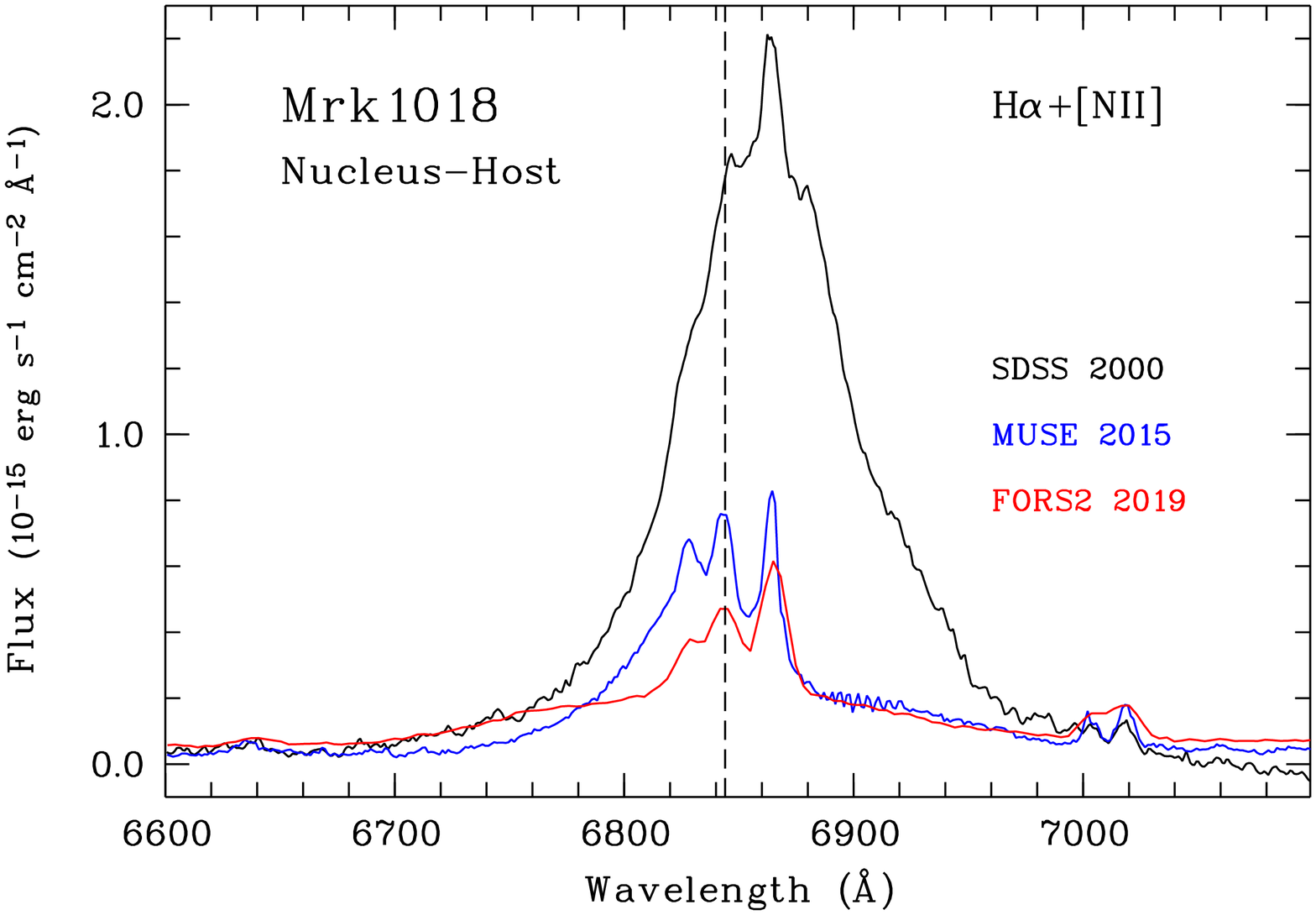}}
\caption{H$\alpha$ + [NII] emission line profile at different epochs. The contribution of the host galaxy has been subtracted from the spectrum of the nucleus. The SDSS spectrum has been divided by a factor of 1.55, as in Fig.~\ref{fig:spec2}, and shifted down by 0.55 units. The dashed line gives the position of the narrow H$\alpha$ line for $z$ = 0.043.}
\label{fig:spec3}
\end{figure}

We applied the same procedure to remove the host galaxy contribution from the MUSE 2015 data and from the Sloan Digital Sky Survey (SDSS) spectrum observed in 2000, the SDSS spectrum being representative of Mrk~1018 in its type~1 state. The MUSE host galaxy template was built from two off-nucleus spectra that were extracted using the same subslits as those used for the FORS2 data, and scaled so as to minimize the absorption features in the difference spectrum. For the SDSS spectrum, we used the host galaxy template built from the MUSE data. The spectra were calibrated in absolute flux with respect to each other, assuming that the contribution of the host galaxy did not change with time. MUSE spectra -- extracted with a circular aperture of 3\arcsec, like the SDSS spectra, and  with the same rectangular subslit as for the FORS2 spectra -- were used to estimate the slit loss for both the point-source AGN and the extended host galaxy that dominates the MUSE and FORS2 continua. The resulting AGN difference spectra are shown in Fig.~\ref{fig:spec2}. The FORS2 and MUSE AGN spectra appear very similar, with almost nothing left of the strong type~1 continuum seen in the SDSS spectrum; this is in agreement with \citet{2016McElroy}, who subtracted a model fitting the host spectrum. The residual AGN continuum slightly increases toward the red in both the FORS2 and MUSE spectra, a consequence of the fact that the host galaxy template is slightly bluer than the nucleus itself (Fig.~\ref{fig:spec1}).

Figure~\ref{fig:spec3} shows the evolution of the H$\alpha$ emission line profile. The broad emission line appears similar in 2019 and in 2015, with some variations already reported by \citet{2017Krumpe} from spectra obtained in 2016-2017. While the MUSE and FORS2 spectra superpose almost perfectly in the red wing, emission in the blue wing at 6800 \AA\ is fainter in 2019 than in 2015, as observed in 2016 by \citet{2017Krumpe}. On the other hand, the 6700-6780~\AA\ blue wing is stronger in 2019 than in 2015; it is not detected in the MUSE spectrum but is seen clearly in the SDSS spectrum. Together with the red wing, it constitutes a very broad, nearly symmetric, emission component with a full width at zero intensity FWZI $\simeq$ 13000 km~s$^{-1}$. In order to fit the SDSS and MUSE H$\alpha$+[NII] line profiles with Gaussians that represent the narrow and broad emission lines, two broad emission lines with full widths at half maximum FWHMs $\simeq$ 3500 and 8700~km~s$^{-1}$ (SDSS) or 2500 and 6700~km~s$^{-1}$ (MUSE) are necessary. The need for two broad lines was already noticed by \citet{2018Kim}. On the contrary, a single broad emission line of FWHM $\simeq$ 7200~km~s$^{-1}$, centered on the narrow H$\alpha$ line, is sufficient to fit the FORS2 H$\alpha$+[NII] line profile (in addition to the narrow lines). This suggests that the $\sim$3000~km~s$^{-1}$-broad line disappeared in 2019. Although the line is fainter, a similar behavior is observed in the H$\beta$ broad emission line profile (Fig.~\ref{fig:spec2}), which can also be fit with a single, centered 7200~km~s$^{-1}$-wide broad emission line.

\subsection{Polarization properties}

\begin{figure}[t]
\centering
\resizebox{\hsize}{!}{\includegraphics*{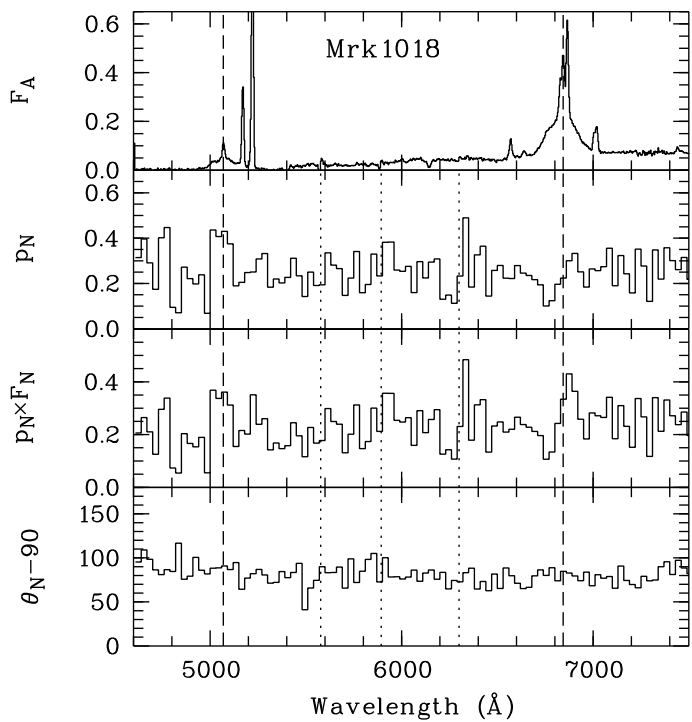}}
\caption{Polarization of the nucleus of Mrk~1018 in the H$\beta$--H$\alpha$ spectral region. The polarization degree $p_{\rm \scriptscriptstyle N}(\lambda)$ is given in percent, the polarized flux density $p_{\rm \scriptscriptstyle N}(\lambda) \times F_{\rm \scriptscriptstyle N}(\lambda)$ in 10$^{-17}$~erg~s$^{-1}$~cm$^{-2}$~\AA$^{-1}$, and the polarization angle $\theta_{\rm \scriptscriptstyle N}(\lambda)$ in degree. Contrary to the polarization degree, the polarized flux density does not depend on the dilution by unpolarized light. These spectra were computed from the Stokes $q_{\rm \scriptscriptstyle N}(\lambda)$ and $u_{\rm \scriptscriptstyle N}(\lambda)$ binned over 30 pixels to increase the signal-to-noise ratio. The AGN direct spectrum $F_{\rm \scriptscriptstyle A} = F_{\rm \scriptscriptstyle N} - F_{\rm \scriptscriptstyle Host}$ (Fig.~\ref{fig:spec1}) is shown unbinned. Vertical dashed lines indicate the position of the H$\beta$ and H$\alpha$ narrow lines, and dotted lines indicate the position of major sky lines.}
\label{fig:spol1}
\end{figure}

Spectropolarimetry of the nucleus of  Mrk~1018 is shown in Fig.~\ref{fig:spol1} for the  H$\beta$--H$\alpha$ spectral range.  Table~\ref{tab:pola} provides integrated polarization measurements: the normalized Stokes parameters $q$ and $u$, the polarization degree $p$, its error $\sigma_p \simeq  \sigma_q \simeq \sigma_u$, the polarization position angle $\theta$, and its error $\sigma_{\theta} = 28.65\degr \times \sigma_p / p$. The polarization of the nucleus continuum was integrated blueward (B) and redward (R) of H$\alpha$ over the wavelength ranges 5250$-$6650~\AA\ and  7050$-$8750~\AA , respectively. The polarizations of the H$\beta$ and H$\alpha$ broad emission lines  were integrated over the wavelength ranges 4995$-$5145~\AA\  and 6700$-$6990~\AA , respectively. The measurements for the continuum polarization were also done for the host galaxy windows defined in Fig.~\ref{fig:ima}.

\begin{table}[t]
\caption{Polarization measurements.}
\label{tab:pola}
\centering
\begin{tabular}{lcrrrrrr}
\hline\hline
 &  & $q$ \ \ & $u$ \ \ &  $p$ \ \ & $\sigma_p$ \ &  $\theta$ \   & $\sigma_{\theta}$   \\
       &       & (\%) & (\%) & (\%)  & (\%)       &  ($\degr$) & ($\degr$) \\
\hline 
Nucleus       & B         &  0.22  & -0.08 & 0.23 & 0.01 & 170 & 1 \\
              & R         &  0.24  & -0.05 & 0.25 & 0.01 & 174 & 1 \\
              & H$\beta$  &  0.37  & -0.04 & 0.37 & 0.04 & 177 & 3 \\
              & H$\alpha$ &  0.20  & -0.07 & 0.21 & 0.02 & 170 & 3 \\
Host North    & B         & -0.07  & -0.21 & 0.22 & 0.03 & 126 & 4 \\
              & R         &  0.03  & -0.16 & 0.16 & 0.03 & 140 & 5 \\
Host South    & B         & -0.08  & -0.08 & 0.11 & 0.03 & 112 & 8 \\
              & R         &  0.09  &  0.05 & 0.10 & 0.03 &  15 & 9 \\
\hline
\end{tabular}
\end{table}

The polarization of the continuum measured in the nuclear region is very small ($p$ = 0.23$\pm$0.01\%  and $\theta$ = 170$\pm$1\degr\ in the B spectral range) and identical within uncertainties to the polarization measured in 1986 when Mrk~1018 was in a much brighter type~1 state ($p$ = 0.28$\pm$0.05\%, $\theta$ = 165$\pm$5\degr; \citealt{1989Goodrich}). This suggests that the polarization is dominated by interstellar dust polarization (dichroic extinction) in our Galaxy and/or in the Mrk~1018 host galaxy.  The polarization degree of the continuum of the host galaxy is smaller (southern part) or comparable (northern part) to the polarization degree measured in the nucleus continuum. This suggests that the polarization of the nuclear region mostly originates from the dust dichroic extinction of stellar light in the host galaxy (with a possible contribution from dust scattering) rather than in our Galaxy.  The polarization of stars\footnote{HD13043 has $p$=0.013$\pm$0.010\%, and HD12641 has $p$=0.034$\pm$0.020\%.} from the \citet{2000Heiles} catalog with angular distances about 30\arcmin\ from Mrk~1018 suggests that the Galactic interstellar polarization is indeed very small along this line of sight. The fact that the polarization position angle of the nucleus is close to the position angle of the major axis of the host galaxy (PA $\simeq$ 0\degr, \citealt{1989Paturel}) also supports a significant contribution from the host galaxy. It is nevertheless important to stress that the polarization measured in the off-centered windows does not represent the host galaxy contribution in the nucleus window, which is likely smaller for symmetry reasons. Indeed, polarization vectors are usually tangential to the galaxy spiral structure \citep[e.g.,][]{1991Scarrott,1992Draper} so that one may expect the polarization degree integrated in the off-centered regions to be higher than in the central region, and with a different polarization angle.

On the other hand, the polarization changes seen in the broad H$\alpha$ line (Fig.~\ref{fig:spol1}) indicate that not all polarization originates from polarized stellar light in the host galaxy, that is, that some originates from the AGN itself; this is most probably a contribution from scattering in the AGN core, which is the case in many type~1 Seyfert galaxies \citep[e.g.][]{2002Smith}.  The fact that the AGN polarization does not increase in the faint state suggests that the AGN dimming and change of look do not result from a dusty cloud that hides the direct continuum source and the broad emission line region, similarly to what is found in changing-look quasars \citep{2017Hutsemekers,2019Hutsemekers}. To fix the ideas, let us write the normalized Stokes parameter $q$ at epochs 1 and 2,
\begin{eqnarray}
  q_{\rm \scriptscriptstyle N1} & = & \alpha _{\rm \scriptscriptstyle 1} \; q_{\rm \scriptscriptstyle A1} \;   + \; (1-\alpha _{\rm \scriptscriptstyle 1}) \; q_{\rm \scriptscriptstyle H} \\
  q_{\rm \scriptscriptstyle N2} & = & \alpha _{\rm \scriptscriptstyle 2} \; q_{\rm \scriptscriptstyle A2} \;   + \; (1-\alpha _{\rm \scriptscriptstyle 2}) \; q_{\rm \scriptscriptstyle H} \; ,
\end{eqnarray}
where  $\alpha = 1/(1+F_{\rm H}/F_{\rm A})$ and {\small N, A, and H} refer to nucleus, AGN, and host galaxy, respectively. The host polarization corresponds to the stellar light polarization integrated over the nucleus subslit. Measurements give $q_{\rm \scriptscriptstyle N1}$ = 0.24\% (in 1986) and  $q_{\rm \scriptscriptstyle N2}$ = 0.22\% (in 2019). Within our aperture, we roughly estimate $\alpha _{\rm \scriptscriptstyle 1} \simeq 0.5$ in the bright state and $\alpha _{\rm \scriptscriptstyle 2} \simeq 0.03$ in the faint state (Figs.~\ref{fig:spec1}~and~\ref{fig:spec2}, and \citealt{2016McElroy}). If the polarized flux ($q \times F$) remains unchanged when the direct flux is strongly dimmed (for example if the direct continuum is occulted by a dusty cloud), we would have $q_{\rm \scriptscriptstyle N2} = q_{\rm \scriptscriptstyle N1} \times (1-\alpha _{\rm \scriptscriptstyle 2})/(1-\alpha _{\rm \scriptscriptstyle 1}) \simeq 2  \, q_{\rm \scriptscriptstyle N1}$, which is not observed. We may therefore assume that the AGN polarized flux does actually change with a time delay between the direct and the polarized light, as modeled in \citet{2020Marin}. However, the polarization angle of the continuum does not flip by 90\degr\ between 1986 and 2019. Such a flip is expected if the AGN polarization dominated by equatorial scattering fades out prior to polarization dominated by polar scattering \citep{2020Marin}. This suggests that some parameters of the model should be revised. In particular, the polarization angle flip could be hidden by the contribution of the host galaxy polarization. However, according to Eq.~2, a negative $q_{\rm \scriptscriptstyle A2}$ (that is, a 90\degr\ flip of the AGN polarization if $q_{\rm \scriptscriptstyle A1}$ is positive) would imply $q_{\rm \scriptscriptstyle H} > q_{\rm \scriptscriptstyle N2}$ to keep $q_{\rm \scriptscriptstyle N2}$ positive (no observed flip), that is, a polarization from the host higher in the central region than in the northern region, which, as previously discussed, seems unlikely.

Although the polarization spectra shown in Fig.~\ref{fig:spol1} are rather noisy (with a few spurious features due to improper subtraction of sky lines), polarization changes across the H$\alpha$ broad emission line are clearly detected. The asymmetry observed in the polarization profile together with the absence of polarization angle rotation could indicate that H$\alpha$ originates in an outflow along the AGN polar axis rather than in a rotating disk \citep{2005Smith,2008Axon}. Alternatively, the dip in the polarization degree precisely corresponds to the blue wing of the profile that significantly brightened between 2015 and 2019 (Fig.~\ref{fig:spec3}). We may thus speculate that the light from this part of the profile is less polarized than the red part due to the time delay between the direct and the polarized light. A more elaborated interpretation of the polarization profile would require a proper subtraction of the polarization from the host galaxy that is not known in the central window. Finally, it is interesting to point out that the polarization signatures predicted for broad lines emitted around supermassive binary black holes, in particular a symmetric double-peaked change in the polarization angle across the line profile \citep{2019Savic,2019Savicb}, are not observed. If confirmed at other epochs, this could rule out the idea that the change of look in Mrk~1018 is due to a recoiling supermassive black hole \citep{2018Kim}.

\section{Conclusions}
\label{sec:conclu}

The AGN at the center of Mrk~1018 was in a faint type 1.5 state in October 2019. The AGN spectrum shows an extremely weak continuum with faint H$\beta$ and H$\alpha$ broad emission lines. Both lines can be fit with a single very broad emission line component of FWHM $\simeq$~7200~km~s$^{-1}$, suggesting that the previously reported 3000~km~s$^{-1}$-wide component has disappeared. If the Balmer lines originate in a Keplerian disk, a decrease in the AGN luminosity will indeed limit the ionization to the central, higher velocity, regions of the disk. On the other hand, the fact that the high velocity blue wing was detected in 2000, was absent in 2015, and was back in 2019 suggests that the Balmer lines are formed in a more complex region that may include a disk wind or a polar outflow. A re-brightening \citep{2017Krumpe} would first raise the blue wing of a line that arises in a polar outflow. 

The continuum polarization of Mrk~1018 is low. Although dominated by polarization from the host galaxy, it shows an intrinsic AGN component that is likely due to scattering. The continuum polarization is not higher in the current faint type~1.5 state than in the past bright type~1 state, confirming that dust obscuration is not the mechanism at the origin of the change of look. The polarization profile of the H$\alpha$ line is asymmetric with no rotation of the polarization angle, possibly revealing line formation in a polar outflow. Alternatively, the polarization profile can be due to the time delay between the direct and the polarized light. Further observations, in particular long-term spectropolarimetric  monitoring, are needed to disentangle these scenarios and better constrain the models. Interestingly, the polarization signatures predicted for broad lines emitted around supermassive binary black holes are not observed.

%

\bibliographystyle{aa}
\bibliography{references}

\end{document}